\documentstyle{article}

\newcommand{\Bd}{B_d^0}
\newcommand{\Bdbar}{\bar B_d^0}
\newcommand{\bd}{\left| B_d^0 \right\rangle}
\newcommand{\bdbar}{\left| \bar B_d^0 \right\rangle}

\newcommand{\KK}{K^0}
\newcommand{\Kbar}{\bar K^0}
\newcommand{\kk}{\left| K^0 \right\rangle}
\newcommand{\kbar}{\left| \bar K^0 \right\rangle}

\newcommand{\kl}{\left| K_L \right\rangle}
\newcommand{\ks}{\left| K_S \right\rangle}

\title{CPT violation and cascade decays}

\author{L.\ Lavoura \\
\small Universidade T\'ecnica de Lisboa \\
\small Centro de F\'\i sica das Interac\c c\~oes Fundamentais \\
\small Instituto Superior T\'ecnico, 1049-001 Lisboa, Portugal}

\date{3 November 1999}

\begin{document}
\maketitle

\begin{abstract}
In  tagged cascade decays of the type
$P^0 (\bar P^0) \rightarrow M^0\!/\!\bar M^0 X \rightarrow f X$
there are two complex CPT-violating parameters:
one in $P^0$--$\bar P^0$ mixing,
the other one in $M^0$--$\bar M^0$ mixing.
I analyze those decays and find that,
in principle,
the former parameter can be extracted
from a careful comparison of their time dependences.
CPT violation in $M^0$--$\bar M^0$ mixing,
on the other hand,
always appears entangled with ratios of decay amplitudes
and cannot be extracted. 
\end{abstract}

\section{Introduction}

The CPT theorem states that any local field theory with Hermitian,
Lorentz-invariant interactions
obeying the spin--statistics connection
is necessarily CPT-invariant \cite{proofs}.
Although the assumptions of this theorem---and thus
the validity of its conclusion---are generally taken for granted,
the question of whether CPT is violated or not
should ultimately be settled by experiment.

The smallness of the mixing
between a neutral meson $P^0$ and its antiparticle $\bar P^0$
(here $P^0$ may be either $K^0$,
$D^0$,
$B^0_d$,
or $B^0_s$)
makes it an ideal setting to look for violations of the discrete symmetries.
Recently,
the CPLEAR Collaboration has presented the results
of its search for T violation \cite{CPLEAR-T}
and CPT violation \cite{CPLEAR-CPT}
in $\KK$--$\Kbar$ mixing.
The OPAL Collaboration has looked for CPT-violating effects
in the $\Bd$--$\Bdbar$ system \cite{OPAL}.
Detailed experiments on these systems
are planned for the $\phi$- and $\Upsilon(4S)$-factories.

It has been pointed out long time ago \cite{enz}
that CPT violation in neutral-meson mixing
can in principle be identified
by observing the time dependence of tagged decays,
{\it i.e.},
by comparing the time dependence of $P^0 \rightarrow f$
with the one of $\bar P^0 \rightarrow f$,
for whatever final state $f$.
The subject has been studied in detail more recently \cite{lavoura}.
The deleterious effect of possible mistags
has also been discussed \cite{previous}.

In this paper I consider the possibility
of probing CPT violation in tagged cascade decays of the type
$P^0 (\bar P^0) \rightarrow M^0\!/\!\bar M^0 X \rightarrow f X$.
There are in this case two CPT-violating parameters,
one in $P^0$--$\bar P^0$ mixing,
the other one in $M^0$--$\bar M^0$ mixing.
On the other hand,
the time dependence to be observed is richer
than the one in non-cascade decays,
since there are two relevant times:
$t$,
the time $P$ mesons oscillate before decaying into $M X$,
and $t^\prime$,
the time $M$ mesons oscillate before decaying into $f$.

Specifically,
I study the important instance of the decay chain
$\Bd (\Bdbar) \rightarrow J/\psi\, \KK\!/\!\Kbar
\rightarrow J/\psi\, f$,
in which $f$ is the final state into which the neutral kaon decays
($f$ may for instance be $\pi^+ \pi^-$ or $\pi^- e^+ \nu_e$).
In the standard model,
the analysis of this decay \cite{cascade} is based on the fact that
the process $\Bd \rightarrow J/\psi\, \Kbar$
does not exist at tree level.
However,
new-physics (and higher-order) effects may alter this situation,
and one cannot neglect those effects
when considering such a radical possibility as CPT violation.
For this reason,
and also in order to obtain a more general treatment,
valid for any cascade decay,
I take into account non-vanishing amplitudes for
$\Bd \rightarrow J/\psi\, \Kbar$ and for
$\Bdbar \rightarrow J/\psi\, \KK$.
Thus,
my conclusions also hold for cascade decays
of the types $B \rightarrow D$ or $D \rightarrow K$,
although in the latter cases complete measurements
will probably be almost impossible to carry out in practice.

The cascade decay referred to above
is often treated in terms of the long- and short-lived neutral kaons
as $\Bd (\Bdbar) \rightarrow J/\psi\, K_S\!/\!K_L$.
However,
such a treatment is problematic,
because---unless CP is conserved---$K_S$ and $K_L$ are not physical particles,
and they cannot be unequivocally identified experimentally.
Indeed,
$K_S$ and $K_L$ are just phenomenological constructs,
which allow one to better visualize the time evolution
of a neutral-kaon beam.
Unless CP is conserved,
there is no decay product of a neutral kaon
which unmistakeably identifies its predecessor
as being either a $K_S$ or a $K_L$.
When making a precise analysis
such as the one required to probe CPT invariance,
one cannot use questionable procedures like truncating the $K_L$-component
(or the $K_S$-component)
of the time-evolution operator,
or assuming that $K_L$ (or $K_S$) does not decay into a certain final state.

I assume that the CP- and CPT-violating parameters
are Lorentz-invariant,
{\it i.e.},
that they do not depend on the four-momentum of the neutral mesons.
This assumption is hidden,
but present,
in the overwhelming majority of previous phenomenological analyses,
as well as in all previous experimental work on CPT violation.
My conclusions are strictly valid only within this scenario,
which however does not apply for theories of CPT violation
which are not Lorentz-invariant,
like the one recently developed
by Kosteleck\'y and collaborators \cite{kostelecky 2}.

\section{Mixing in the $B$ system}

The time evolution of
\begin{equation}
| \psi (t) \rangle = \psi_1 (t) \bd + \psi_2 (t) \bdbar
\label{the main state}
\end{equation}
is given by
\begin{equation}
i \frac{d}{dt}
\left( \begin{array}{c} \psi_1 (t) \\ \psi_2 (t) \end{array} \right) =
\left( \begin{array}{cc} R_{11} & R_{12} \\ R_{21} & R_{22} \end{array} \right)
\left( \begin{array}{c} \psi_1 (t) \\ \psi_2 (t) \end{array} \right).
\label{the main evolution}
\end{equation}
(The matrix $R$ is usually written $R = M - i \Gamma / 2$,
with $M$ and $\Gamma$ Hermitian.)
There is CPT and CP violation in the mixing if $R_{11} \neq R_{22}$.
There is T and CP violation in the mixing
if $\left| R_{12} \right| \neq \left| R_{21} \right|$.

The eigenvalues of $R$ are denoted $\mu_a$ and $\mu_b$.
Their sum is equal to the trace of $R$:
\begin{equation}
\mu_a + \mu_b = R_{11} + R_{22}.
\label{trace}
\end{equation}
The right-eigenvectors of $R$
corresponding to the eigenvalues $\mu_a$ and $\mu_b$
are $(p_a, q_a)^T$ and $(p_b, -q_b)^T$,
respectively:
\begin{eqnarray}
\left( \begin{array}{cc} R_{11} & R_{12} \\ R_{21} & R_{22} \end{array} \right)
\left( \begin{array}{c} p_a \\ q_a \end{array} \right)
&=& \mu_a
\left( \begin{array}{c} p_a \\ q_a \end{array} \right),
\nonumber\\
\left( \begin{array}{cc} R_{11} & R_{12} \\ R_{21} & R_{22} \end{array} \right)
\left( \begin{array}{c} p_b \\ - q_b \end{array} \right)
&=& \mu_b
\left( \begin{array}{c} p_b \\ - q_b \end{array} \right).
\label{eigenvectors}
\end{eqnarray}
We do not need to make any assumption about either the normalization
or the relative phase of the eigenvectors.
Equations~(\ref{eigenvectors}) may be written
\begin{eqnarray}
\frac{q_a}{p_a}
&=& \frac{\mu_a - R_{11}}{R_{12}}
= \frac{R_{21}}{\mu_a - R_{22}},
\nonumber\\
\frac{q_b}{p_b}
&=& \frac{R_{11} - \mu_b}{R_{12}}
= \frac{R_{21}}{R_{22} - \mu_b}.
\label{eigenvectors, explicit}
\end{eqnarray}
Equations~(\ref{trace}) and (\ref{eigenvectors, explicit}) imply
\begin{equation}
\theta \equiv \frac{q_a / p_a - q_b / p_b}{q_a / p_a + q_b / p_b}
= \frac{R_{22} - R_{11}}{\mu_a - \mu_b}.
\label{theta}
\end{equation}
The dimensionless complex number $\theta$
parametrizes the violation of CPT in the mixing.

It is convenient to introduce
\begin{equation}
\frac{q}{p} \equiv \sqrt{\frac{q_a q_b}{p_a p_b}}
= \sqrt{\frac{R_{21}}{R_{12}}}.
\label{ratio}
\end{equation}
From Eqs.~(\ref{theta}) and (\ref{ratio}) it follows that
\begin{equation}
\sqrt{1 - \theta^2} = \frac{2 q / p}{q_a / p_a + q_b / p_b}.
\label{square root}
\end{equation}
If CPT violation is absent from the mixing,
then $q/p = q_a / p_a = q_b / p_b$ and $\sqrt{1 - \theta^2} = 1$.
In that case one only needs to use $q/p$.

It follows from Eqs.~(\ref{the main state}),
(\ref{the main evolution}),
and (\ref{eigenvectors}) that the states
\begin{eqnarray}
\left| B_a \right\rangle &=& p_a \bd + q_a \bdbar,
\nonumber\\
\left| B_b \right\rangle &=& p_b \bd - q_b \bdbar
\label{ph and pl}
\end{eqnarray}
evolve in time as
\begin{eqnarray}
\left| B_a (t) \right\rangle &=& e^{- i \mu_a t} \left| B_a \right\rangle,
\nonumber\\
\left| B_b (t) \right\rangle &=& e^{- i \mu_b t} \left| B_b \right\rangle.
\label{primary evolution}
\end{eqnarray}
Let us consider a neutral meson which is tagged as a $\Bd$ ($\Bdbar$)
at time $t=0$.
Using Eqs.~(\ref{theta})--(\ref{primary evolution}),
one finds that this state is given at time $t$ by
\begin{eqnarray}
\left| \Bd (t) \right\rangle &=&
\left[ g_+ (t) - \theta g_- (t) \right] \bd
+ \frac{q}{p} \sqrt{1 - \theta^2} g_- (t) \bdbar,
\nonumber\\
\left| \Bdbar (t) \right\rangle &=&
\left[ g_+ (t) + \theta g_- (t) \right] \bdbar
+ \frac{p}{q} \sqrt{1 - \theta^2} g_- (t) \bd,
\label{evolution}
\end{eqnarray}
respectively.
Here,
\begin{equation}
g_\pm (t) \equiv {\textstyle \frac{1}{2}}
\left( e^{- i \mu_a t} \pm e^{- i \mu_b t} \right).
\end{equation}

\section{Mixing in the $K$ system}

Contrary to most authors,
we treat mixing in the $K$ system in an exactly analogous fashion
to mixing in the $B$ system \cite{livro}.
We do not introduce the usual parameters $\epsilon$ and $\delta$,
which are in general defined in a non-rephasing-invariant way,
and have an awkward interpretation \cite{bb}.

The time evolution of
\begin{equation}
| \varphi (t^\prime) \rangle
= \varphi_1 (t^\prime) \kk + \varphi_2 (t^\prime) \kbar
\label{the main state prime}
\end{equation}
is given by
\begin{equation}
i \frac{d}{dt^\prime}
\left( \begin{array}{c}
\varphi_1 (t^\prime) \\ \varphi_2 (t^\prime)
\end{array} \right)
= \left( \begin{array}{cc}
R_{11}^\prime & R_{12}^\prime \\ R_{21}^\prime & R_{22}^\prime
\end{array} \right)
\left( \begin{array}{c}
\varphi_1 (t^\prime) \\ \varphi_2 (t^\prime)
\end{array} \right).
\label{the main evolution prime}
\end{equation}
CPT is violated if $R^\prime_{11} \neq R^\prime_{22}$,
and T is violated
if $\left| R^\prime_{12} \right| \neq \left| R^\prime_{21} \right|$.

The eigenvalues of $R^\prime$ are denoted $\mu_L$ and $\mu_S$.
Their sum is equal to the trace of $R^\prime$:
\begin{equation}
\mu_L + \mu_S = R_{11}^\prime + R_{22}^\prime.
\label{trace prime}
\end{equation}
The right-eigenvectors of $R^\prime$
corresponding to the eigenvalues $\mu_L$ and $\mu_S$
are $(p_L, q_L)^T$ and $(p_S, -q_S)^T$,
respectively:
\begin{eqnarray}
\left( \begin{array}{cc} R_{11}^\prime & R_{12}^\prime \\
R_{21}^\prime & R_{22}^\prime \end{array} \right)
\left( \begin{array}{c} p_L \\ q_L \end{array} \right)
&=& \mu_L
\left( \begin{array}{c} p_L \\ q_L \end{array} \right),
\nonumber\\
\left( \begin{array}{cc} R_{11}^\prime & R_{12}^\prime \\
R_{21}^\prime & R_{22}^\prime \end{array} \right)
\left( \begin{array}{c} p_S \\ - q_S \end{array} \right)
&=& \mu_S
\left( \begin{array}{c} p_S \\ - q_S \end{array} \right).
\label{eigenvectors prime}
\end{eqnarray}
We do not make any assumption about either the normalization
or the relative phase of the eigenvectors.
Equations~(\ref{eigenvectors prime}) may be written
\begin{eqnarray}
\frac{q_L}{p_L}
&=& \frac{\mu_L - R_{11}^\prime}{R_{12}^\prime}
= \frac{R_{21}^\prime}{\mu_L - R_{22}^\prime},
\nonumber\\
\frac{q_S}{p_S}
&=& \frac{R_{11}^\prime - \mu_S}{R_{12}^\prime}
= \frac{R_{21}^\prime}{R_{22}^\prime - \mu_S}.
\label{eigenvectors, explicit prime}
\end{eqnarray}
Equations~(\ref{trace prime}) and (\ref{eigenvectors, explicit prime})
imply
\begin{equation}
\theta^\prime \equiv
\frac{q_L/p_L - q_S/p_S}{q_L/p_L + q_S/p_S}
= \frac{R_{22}^\prime - R_{11}^\prime}{\mu_L - \mu_S}.
\label{theta prime}
\end{equation}
The parameter $\theta^\prime$ violates CPT and CP.

It is convenient to introduce
\begin{equation}
\frac{q^\prime}{p^\prime} \equiv \sqrt{\frac{q_L q_S}{p_L p_S}}
= \sqrt{\frac{R_{21}^\prime}{R_{12}^\prime}}.
\label{ratio prime}
\end{equation}
If $\left| q^\prime / p^\prime \right| \neq 1$ then T and CP are violated.
The phase of $q^\prime / p^\prime$ is physically meaningless.

It follows from Eqs.~(\ref{the main state prime}),
(\ref{the main evolution prime}),
and (\ref{eigenvectors prime}) that the states
\begin{eqnarray}
\kl &=& p_L \kk + q_L \kbar,
\nonumber\\
\ks &=& p_S \kk - q_S \kbar
\label{kl and ks}
\end{eqnarray}
evolve in time as
\begin{eqnarray}
\left| K_L \! \left( t^\prime \right) \right\rangle &=&
e^{- i \mu_L t^\prime} \kl,
\nonumber\\
\left| K_S \! \left( t^\prime \right) \right\rangle &=&
e^{- i \mu_S t^\prime} \ks.
\label{primary evolution prime}
\end{eqnarray}
We may invert Eqs.~(\ref{kl and ks})
and use Eqs.~(\ref{primary evolution prime}) to find
\begin{eqnarray}
\left| \KK \! \left( t^\prime \right) \right\rangle
&=& \frac{q_L}{p_L q_S + p_S q_L}
\left( e^{- i \mu_S t^\prime} \ks
+ \frac{q_S}{q_L}\, e^{- i \mu_L t^\prime}\kl \right),
\nonumber\\
\left| \Kbar \! \left( t^\prime \right) \right\rangle
&=& \frac{q_L}{p_L q_S + p_S q_L}
\left( - \frac{p^\prime}{q^\prime} \right)
\left( \sqrt{\frac{1 - \theta^\prime}{1 + \theta^\prime}}
e^{- i \mu_S t^\prime} \ks \right.
\nonumber\\
& & \left. - \sqrt{\frac{1 + \theta^\prime}{1 - \theta^\prime}}\,
\frac{q_S}{q_L}\, e^{- i \mu_L t^\prime} \kl \right).
\label{ks inverted}
\end{eqnarray}

Consider the decay of a neutral kaon into the state $f$.
Inspired by Eqs.~(\ref{ks inverted}),
we define
\begin{equation}
\eta_f \equiv \frac{q_S}{q_L}\, \frac{\langle f | T \kl}{\langle f | T \ks}.
\label{eta definition}
\end{equation}
The prefactor $q_S / q_L$
(or some other analogous prefactor)
is necessary \cite{kayser} in order to account for
the possibly different normalizations,
and phases,
of $\kl$ and $\ks$.
In particular,
$\eta_f$ is invariant under a rephasing of $\kl$ and $\ks$.
In general,
one may make the convention $q_S = q_L$ for the normalizations
and phases of $\kl$ and $\ks$;
one then obtains $\eta_f = \langle f | T \kl / \langle f | T \ks$,
the definition that one normally finds in the literature.
However,
making that convention is not mandatory.

\section{Cascade decays}

We consider the decays of the neutral $B$ mesons
to $J/\psi$ and a neutral kaon.
We normalize the four relevant decay amplitudes
by $\langle J/\psi \KK | T | \Bd \rangle$.
We thus define the three parameters
\begin{eqnarray}
y &\equiv&
- \frac{\langle J/\psi \Kbar | T | \Bd \rangle}
{\langle J/\psi \KK | T | \Bd \rangle}\,
\frac{p^\prime}{q^\prime},
\label{y definition}\\
\bar y &\equiv&
\frac{q}{p}\,
\frac{\langle J/\psi \KK | T | \Bdbar \rangle}
{\langle J/\psi \KK | T | \Bd \rangle},
\label{y-bar definition}\\
\lambda &\equiv&
- \frac{q}{p}\,
\frac{\langle J/\psi \Kbar | T | \Bdbar \rangle}
{\langle J/\psi \KK | T | \Bd \rangle}\,
\frac{p^\prime}{q^\prime}.
\label{lambda definition}
\end{eqnarray}
These parameters are rephasing-invariant,
{\it i.e.},
they are invariant under an arbitrary rephasing of the kets $\bd$,
$\bdbar$,
$\kk$,
and $\kbar$.
In the standard model $\lambda = \exp \left( - 2 i \beta \right)$
has modulus 1.
On the other hand,
$y$ and $\bar y$ are expected to be very small
and one usually neglects them.

As a matter of fact,
$\bar y$ is like the parameter $\lambda_f$,
with $f = J/\psi \KK$,
which is usually introduced in order to parametrize the interference
between the mixing of the $B$ mesons and their decay into a final state $f$.
On the other hand,
$y$ is one of the parameters $\xi_i$,
with $i = \Bd$,
which were recently introduced by Silva and collaborators \cite{silva}
in order to parametrize the interference between
the decay into a neutral-meson (in this case, the $K$) system
and the mixing in that system.
The parameter $\lambda$ is a more complicated entity,
since it simultaneously involves mixing in the $B$ system,
mixing in the $K$ system,
and the decays from the $B$ into the $K$ system.
However,
it is precisely $\lambda$ which is usually significant,
because the amplitudes $\langle J/\psi \Kbar | T | \Bd \rangle$
and $\langle J/\psi \KK | T | \Bdbar \rangle$
are generally assumed to be negligible.

We consider the experiment in which a tagged $\Bd$ evolves for a time $t$,
then it decays into $J/\psi$ and a neutral kaon.
Afterwards,
the neutral kaon evolves for a time $t^\prime$
and finally decays into $f$.
The amplitude for this process is
\begin{eqnarray}
& & \langle f | T | \KK (t^\prime) \rangle \,
\langle J/\psi \KK | T | \Bd (t) \rangle
+
\langle f | T | \Kbar (t^\prime) \rangle \,
\langle J/\psi \Kbar | T | \Bd (t) \rangle
\nonumber\\
&\propto& e^{- i \mu_S t^\prime}
\left( e^{- i \mu_a t} + R e^{- i \mu_b t} \right)
+ Q \eta_f e^{- i \mu_L t^\prime}
\left( e^{- i \mu_a t} + S e^{- i \mu_b t} \right).
\label{ABC definition}
\end{eqnarray}

We next consider the analogous experiment with an initial $\Bdbar$.
The amplitude is
\begin{eqnarray}
& & \langle f | T | \KK (t^\prime) \rangle\,
\langle J/\psi \KK | T | \Bdbar (t) \rangle
+
\langle f | T | \Kbar (t^\prime) \rangle\,
\langle J/\psi \Kbar | T | \Bdbar (t) \rangle
\nonumber\\
&\propto& e^{- i \mu_S t^\prime}
\left( e^{- i \mu_a t} + \bar R e^{- i \mu_b t} \right)
+ \bar Q \eta_f e^{- i \mu_L t^\prime}
\left( e^{- i \mu_a t} + \bar S e^{- i \mu_b t} \right).
\label{ABC-bar definition}
\end{eqnarray}

The parameters $R, S, \ldots, \bar Q$ defined by Eqs.~(\ref{ABC definition})
and (\ref{ABC-bar definition})
are independent of the final state $f$.
In principle one may measure $R, S, \bar R$ and $\bar S$
by observing the dependence on $t$ and $t^\prime$ of the cascade decay.
On the other hand,
as $Q$ and $\bar Q$ appear multiplied by $\eta_f$,
one cannot really determine them,
but only their ratio.
Upon a simple computation one finds
\begin{eqnarray}
R &=& \frac
{\left( 1 + \theta \right) \left( 1 + y
\sqrt{\frac{1 - \theta^\prime}{1 + \theta^\prime}}
\right) - \sqrt{1 - \theta^2} \left( \bar y + \lambda
\sqrt{\frac{1 - \theta^\prime}{1 + \theta^\prime}}
\right)}
{\left( 1 - \theta \right) \left( 1 + y
\sqrt{\frac{1 - \theta^\prime}{1 + \theta^\prime}}
\right) + \sqrt{1 - \theta^2} \left( \bar y + \lambda
\sqrt{\frac{1 - \theta^\prime}{1 + \theta^\prime}}
\right)},
\nonumber\\
S &=& R \left( \lambda \rightarrow - \lambda, y \rightarrow - y,
\theta^\prime \rightarrow - \theta^\prime \right),
\label{exact expressions}
\end{eqnarray}
and
\begin{eqnarray}
\bar R &=& - \frac{1 - \theta}{1 + \theta}\, R,
\label{equacaoR}\\
\bar S &=& - \frac{1 - \theta}{1 + \theta}\, S,
\label{equacaoS}\\
\bar Q &=& Q.
\label{equacaoQ}
\end{eqnarray}

For better insight,
we may use the first-order approximation for small parameters,
{\it i.e.},
the approximation of neglecting all products of $\theta$,
$\theta^\prime$,
$y$,
and $\bar y$.
One then obtains
\begin{eqnarray}
R &\approx& \frac{1 - \lambda}{1 + \lambda} \left[ 1 +
\frac{2}{1 - \lambda^2} \left( \theta - x \right) \right],
\nonumber\\
S &\approx& \frac{1 + \lambda}{1 - \lambda} \left[ 1 +
\frac{2}{1 - \lambda^2} \left( \theta - x \right) \right],
\label{approximate expressions}
\end{eqnarray}
where
\begin{equation}
x \equiv \bar y - \lambda \theta^\prime - \lambda y.
\label{x definition}
\end{equation}

\section{Identification of CPT violation}

From Eqs.~(\ref{approximate expressions}) and (\ref{x definition})
one concludes that there is no way of measuring $\theta^\prime$,
or even of asserting whether $\theta^\prime$ vanishes or not,
from the observation of the tagged cascade decays.
Indeed,
$\theta^\prime$ always appears entangled with $\lambda$,
$y$,
and $\bar y$.
Now,
even if one might expect,
based on the standard model,
$y$ and $\bar y$ to be vanishingly small,
one can never assert that fact with certainty.
When considering the possibility of CPT violation,
which would be an effect dramatically non-standard,
one cannot start from the assumption
that the expectations of the standard model for some parameters are correct
\cite{previous}.

On the other hand,
one gathers from Eqs.~(\ref{equacaoR}) and (\ref{equacaoS})
that CPT violation in $B$ mixing (the parameter $\theta$)
can in principle be determined either from the comparison of $R$ and $\bar R$,
or from the comparison of $S$ and $\bar S$.
Indeed,
$\bar R \neq - R$ and $\bar S \neq - S$
unequivocally indicate the presence of CPT violation
in the mixing of the $B$ mesons.
One can in principle measure CPT violation in $B$-meson mixing
by observation of the time dependence of the tagged cascade decays.
This conclusion goes in line with previous work \cite{lavoura,previous}.

From Eq.~(\ref{equacaoQ}) one concludes that
$\bar Q / Q = 1$ remains valid in the CPT-violating case.
This relation may serve as a test either of our theoretical understanding,
or of experiment.

CPT violation by the parameter $\theta$ can be identified
even if one omits measuring the time $t^\prime$.
Indeed,
by integrating Eq.~(\ref{ABC definition}) one obtains
\begin{eqnarray}
\int_0^{+ \infty} \!\! dt^\prime
\left| \langle f | T | \KK (t^\prime) \rangle \,
\langle J/\psi \KK | T | \Bd (t) \rangle \right. & &
\nonumber\\
\left. + \langle f | T | \Kbar (t^\prime) \rangle \,
\langle J/\psi \Kbar | T | \Bd (t) \rangle \right|^2
&\propto& e^{- \Gamma_a t} + C e^{- \Gamma_b t}
\nonumber\\
& & + 2 e^{- (\Gamma_a + \Gamma_b) t/2}
\nonumber\\
& & \times \mbox{Re}
\left[ e^{i (m_a - m_b) t} F \right],
\label{CF definition}
\end{eqnarray}
where I have used the notation $m_{a,b} \equiv \mbox{Re}\, \mu_{a,b}$
and $\Gamma_{a,b} \equiv - 2 \mbox{Im}\, \mu_{a,b}$.
Concerning Eq.~(\ref{CF definition}) it is worthwhile to point out that,
if $R \neq S$,
then $|F|^2 < C$;
this differs from what happens in non-cascade decays of the $B$ mesons,
where $|F|^2 = C$.
Similarly to Eq.~(\ref{CF definition}),
\begin{eqnarray}
\int_0^{+ \infty} \!\! dt^\prime
\left| \langle f | T | \KK (t^\prime) \rangle \,
\langle J/\psi \KK | T | \Bdbar (t) \rangle \right. & &
\nonumber\\
\left. + \langle f | T | \Kbar (t^\prime) \rangle \,
\langle J/\psi \Kbar | T | \Bdbar (t) \rangle \right|^2
&\propto& e^{- \Gamma_a t} + \bar C e^{- \Gamma_b t}
\nonumber\\
& & + 2 e^{- (\Gamma_a + \Gamma_b) t/2}
\nonumber\\
& & \times \mbox{Re}
\left[ e^{i (m_a - m_b) t} \bar F \right].
\label{CF-bar definition}
\end{eqnarray}
Now,
one easily checks that
\begin{eqnarray}
\frac{\bar F}{F} &=& - \frac{1 - \theta}{1 + \theta},
\nonumber\\
\frac{\bar C}{C} &=& \left| \frac{1 - \theta}{1 + \theta} \right|^2.
\label{relacoes theta integradas}
\end{eqnarray}
Thus,
$\theta$ can be measured if one compares the $t$ dependence
of the tagged decays of $\Bd$ and of $\Bdbar$,
even if one integrates over $t^\prime$.

\section{Conclusions}

I studied in this work the possibility of probing CPT invariance
by observation of the time dependence of cascade decays of the type
$P^0 (\bar P^0) \rightarrow M^0\!/\!\bar M^0 X \rightarrow f X$.
There are in this case two CPT-violating parameters:
$\theta$ in $P^0$--$\bar P^0$ mixing,
and $\theta^\prime$ in $M^0$--$\bar M^0$ mixing.
There are also two times:
$t$ for $P^0$--$\bar P^0$ oscillations,
and $t^\prime$ for $M^0$--$\bar M^0$ oscillations.

I found that it is possible to determine $\theta$
by comparing the $t$ dependence
of the cascade decays of tagged $P^0$ and tagged $\bar P^0$.
Indeed,
$\theta$ is computed in much the same way as
from the time dependence of non-cascade decays \cite{lavoura}.
One should not forget,
however,
that mistags can ruin the prospects of such measurements \cite{previous}.

On the other hand,
$\theta^\prime$ cannot be determined,
since it always arises
together with some undetermined ratios of decay amplitudes.
In order to determine $\theta^\prime$
(and $\eta_f$ for that matter)
one can have recourse to the tagged decays of the $M$ mesons themselves,
instead of cascade decays
in which $M$ mesons only arise as an intermediate step.

\end{document}